\documentclass[aps,twocolumn,groupedaddress,showpacs,floatfix,amsmath]{revtex4}
\usepackage{tabularx}
\usepackage{graphicx}
\usepackage{dcolumn}
\usepackage{bm}
\usepackage{color}

\definecolor{Remarks}{rgb}{1,0.3,0.3}

\newcommand{\1}{$\bf{(a)}$}
\newcommand{\2}{$\bf{(b)}$}

\newcommand{\be}{\begin{equation}}
\newcommand{\ee}{\end{equation}}
\newcommand{\bea}{\begin{eqnarray}}
\newcommand{\eea}{\end{eqnarray}}
\newcommand{\etal}{{\it{et~al.\,}}}

\begin{document}

\title{Fixed-node diffusion Monte Carlo study of the structures of m-benzyne}
\author{W.~A.~Al-Saidi}
\author{C. J. Umrigar}
\affiliation{Laboratory of Atomic and Solid State Physics,
          Cornell University, Ithaca, New York 14853, USA.}

\date{\today}

\begin{abstract}
Diffusion Monte Carlo (DMC) calculations are performed
on the monocyclic and bicyclic forms of m-benzyne, which are the
equilibrium structures at the CCSD(T) and CCSD levels of coupled
cluster theory.
We employed multi-configuration self-consistent field trial wave functions
which are constructed from a carefully selected
8-electrons-in-8-orbitals complete active space [CAS(8,8)], with  CSF
coefficients that are reoptimized in the presence of a Jastrow factor.
The DMC calculations show that the monocyclic structure is lower
in energy than the bicyclic structure by $1.9(2)$~kcal/mole, in excellent
agreement with the best coupled cluster results.
\end{abstract}

\maketitle

\section{Introduction}

Quantum chemistry methods used to study
1,3-Didehydrobenzene, commonly known as meta-benzyne or m-benzyne,
do not agree upon its geometrical structure.
The issue at question is, which of the structures shown in Fig.~I,
the monocyclic \1 or the bicyclic \2, is lower in energy.
These two structures differ mostly in the distance between the
radical centers, C1 and C3;  the single occupied orbitals on carbon atoms 1 and
3 form a bond in \2 which decreases the biradical
nature of \1.

At the time of its synthesis~\cite{MarquardtSanderKraka96},
coupled cluster calculations with
single and double excitations plus a perturbative treatment of triple
excitations CCSD(T), with  a small 6-31G(d,p) basis~\cite{HariharanPople73}
were used in identifying the structure of m-benzyne. The
computed infrared frequencies of the  monocyclic structure, \1, were in
good agreement with the experimental one, thus indicating that this
structure is the stable one.
Subsequent calculations using larger basis sets also supported this
conclusion~\cite{WinklerSander01,Kraka01,SmithCrawfordCremer05}. 

In an
extensive recent study, Smith, Crawford, and Cremer
\cite{SmithCrawfordCremer05} performed coupled cluster CCSD (single
and double excitations), CCSD(T), and CCSDT (single, double and triple
excitations) calculations on m-benzyne. The CCSD and CCSD(T) calculations were done
with 6-31G(d,p)~\cite{HariharanPople73} and cc-pVTZ \cite{Dunning89} basis sets with
reference determinants from restricted Hartree-Fock (RHF),
unrestricted Hartree-Fock (UHF) and Bruekner orbitals. They found
that CCSD favors the bicyclic form \2 as the ground state with a C1C3
bond length $\approx 1.56$~\AA.  On the other hand, CCSD(T) predicted
\1 as the stable structure with a C1C3 bond length $\approx
2.10$~\AA. With the small 6-31G(d,p) basis set and with RHF orbitals, 
CCSDT also favored the structure \1. One of the conclusions of their 
study is that the inclusion of triple excitations (as in CCSD(T) or
CCSDT) is essential to obtain a correct energy ordering of the structures of m-benzyne,
and, inclusion of  $d$- and $f$-type polarization functions in the basis
is needed for a quantitatively correct energy difference.

\begin{figure}[t]
\label{fig.geometry}
\vspace{0.2in}
\includegraphics[width=8.3 cm]{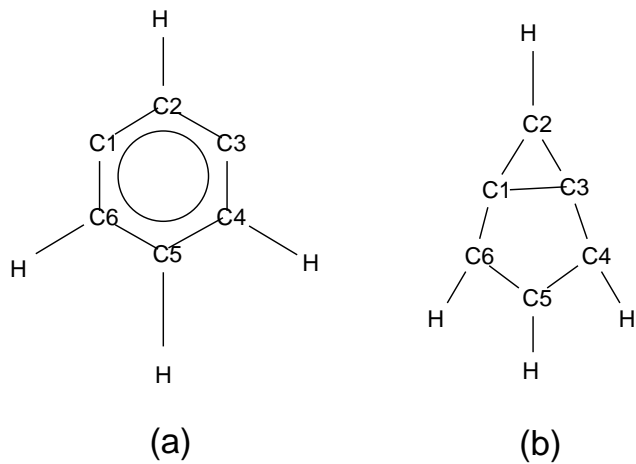}
\caption{Proposed structures of m-benzyne: \1 the monocyclic
  structure and \2 the bicyclic structure.}
\end{figure}

The structural predictions of density functional theory (DFT) are
inconclusive, depending strongly on the choice of the approximate
exchange-correlation functional.  The pure gradient corrected
functionals, Becke-Lee-Yang-Parr (BLYP)~\cite{Becke88,LeeYangParr88}
and Becke-Perdew-Wang-91 (BPW91)~\cite{Becke88,PW91} predict that the
monocyclic form, \1, is stable, whereas, the more complicated
3-parameter B3LYP functional~\cite{b3lyp}, which includes an admixture
of exact exchange, favors the bicyclic structure, \2~\cite{Hess01,
WinklerSander01,Kraka01}.  Since Hartree-Fock theory strongly favors
the bicyclic structure, \2 \cite{WinklerSander01}, it is not
surprising that the hybrid functional would also.  Further, it was
found that the infrared frequencies computed with the B3LYP functional of the two structures agree about
equally well~\cite{Hess01} with the experimental ones, which led the author to
question the previous interpretation ~\cite{MarquardtSanderKraka96}
of the experimental results.

Both coupled cluster and DFT energies are based on single-reference wave functions
and thus their predictions are questionable if m-benzyne has considerable multi-reference
character.  Multi-configuration
self-consistent field (MCSCF) calculations, which do a good job of
describing static correlation but a poor job on dynamic correlation,
indicate that \1 is more stable by a large amount, 12.9
kcal/mole~\cite{Kraka01}. This is to be contrasted with the RHF theory
which predicts that structure \2 is more stable than structure \1 by
about 18 kcal/mole~\cite{WinklerSander01}.  This shows the importance
of a multi-determinant description of m-benzyne; and,  in order to get
reliable energy differences it is essential to have an accurate
description of both static and dynamic correlations.

In this paper we use variational Monte Carlo (VMC) and diffusion
Monte Carlo (DMC) to study this issue.  These methods have the
advantage that both static and dynamic correlation can be described
accurately, provided that well optimized, multi-determinant trial
wave functions are employed.  Further, the results have only a weak
dependence on the size of the basis. Currently, it is still
expensive to optimize the geometry of large molecules using QMC
methods mainly because of their  stochastic nature.  Thus,
we limited our study to the two geometries of \1 and \2 which
are respectively the equilibrium geometries for CCSD(T) and CCSD with
a cc-pVTZ basis~\cite{SmithCrawfordCremer05}.
The geometrical parameters are
reproduced in Table~\ref{tab:geometry} for convenience.

\begin{table}[t]
\caption{
\label{tab:geometry}
Optimized geometrical parameters of structures \1 and \2 with a
cc-pVTZ basis set from Ref.~\cite{SmithCrawfordCremer05}. The first
geometry was optimized using CCSD(T) and the second one with
CCSD. Bond distances are in angstroms and angles are in degrees.}
\begin{ruledtabular}
\begin{tabular}{lcc}
  &  \1 &  \2  \\
\hline
$r_{13}$  &2.026   & 1.551  \\
$r_{23}$  & 1.364  & 1.343  \\
$r_{34}$  & 1.372  & 1.376  \\
$r_{45}$  & 1.398   & 1.404  \\
$r_{2H}$  &1.072  & 1.074  \\
$r_{4H}$  &1.076  & 1.071  \\
$r_{5H}$  &1.080   & 1.078  \\
$\theta_{123}$ & 95.9 & 70.5  \\
$\theta_{345}$& 116.7 & 107.7  \\
$\theta_{456}$& 113.4 & 111.7  \\
$\theta_{34H}$& 120.7 & 126.3  \\
\end{tabular}
\end{ruledtabular}
\end{table}

\section{QMC methods}
\label{qmc_methods}

We employ two QMC methods, variational Monte Carlo (VMC) and diffusion
Monte Carlo (DMC) which have been reviewed before \cite{general_refs1,NigUmr-BOOK-99,general_refs2}.
In both methods the expectation values are obtained
by a Monte Carlo integration and consequently there is great freedom
in the choice of the trial wave function.  VMC yields expectation values
for the trial wave function, whereas in DMC the trial wave function serves as a starting
point for a stochastic projection
onto the ground state.  Although DMC is exact for Bosonic
ground states, for Fermionic systems the projection is performed while imposing
the boundary condition that the nodes of the many-body wave function
are the same as those of the trial states.  Consequently, the DMC
energy is also an upper bound to the true energy and the fixed-node error
depends crucially on the quality of the trial wave function.
In practice, we use an accelerated Metropolis algorithm with a
 very small auto-correlation time~\cite{Umrigar93} and a DMC algorithm~\cite{UNR93}, that takes into account
the singularities in the time-evolution operator and allows one to use
large time steps while still having an acceptably small time-step error.
All the QMC calculations are performed using the program package CHAMP~\cite{Cha-PROG-XX}.

The trial wave function is of the standard
Jastrow-Slater form consisting of a Jastrow factor $J({\bf R}) =
e^{f({\bf R })}$ times a linear combination of determinants.
Specifically,
\be
\Psi_{T}( {\bf R}  ) = J({\bf R}) \sum_{i } c_{i}\, C_i({\bf R})
\ee
where ${\bf{R}} = ( {\bf r}_1, {\bf r}_2, \ldots, {\bf r}_N ) $ is a
$3N$ dimensional vector, ${\bf r}_k$ is the position vector of the kth electron,
$C_i({\bf R})$ is the ith configuration state
function (CSF), which is a symmetry-adapted linear combination of Slater
determinants, and $c_i$ are the CSF linear expansion coefficients.  The Jastrow part
includes explicitly  electron-nucleus, electron-electron, and
electron-electron-nucleus correlation functions \cite{GucluJeonUmrigarJain05}.
The statistical and the systematic errors in the VMC and DMC energies
depend on the quality of the trial wave function.  The Jastrow and CSF variational
parameters are optimized using recently developed energy minimization
methods~\cite{UmrigarFilippi05,UmrigarToulouseFilippiSorellaHennig07,ToulouseUmrigar07}.

To eliminate the core electrons, we employ the energy-consistent scalar-relativistic Hartree-Fock
pseudopotentials of Burkatzki, Filippi and Dolg~\cite{BurkatzkiFilippiDolg07}
and the accompanying basis sets.  For the DMC calculations, we use the standard
locality approximation which~\cite{MitasShirleyCeperley91}
introduces another small systematic error.

\section{Convergence Tests}

Before performing the final calculations, we studied the convergence
of the energy with respect to a) the size of the basis, b) the cutoff
used for selecting which CSFs are included in the expansion, and c) the Trotter time-step used to evolve the system.
All of our DFT and MCSCF calculations are performed using GAMESS~\cite{GAMESS}.

\subsubsection{Basis set convergence}

The VMC and DMC energies depend on the quality of the basis used to
expand the orbitals in the determinants.  One
of the advantages of both QMC methods, particularly DMC, is that the dependence on the basis
size is much weaker than for other quantum chemistry methods.

\begin{figure}[t]
\vspace{0.2in}
\includegraphics[width=8.5cm,clip]{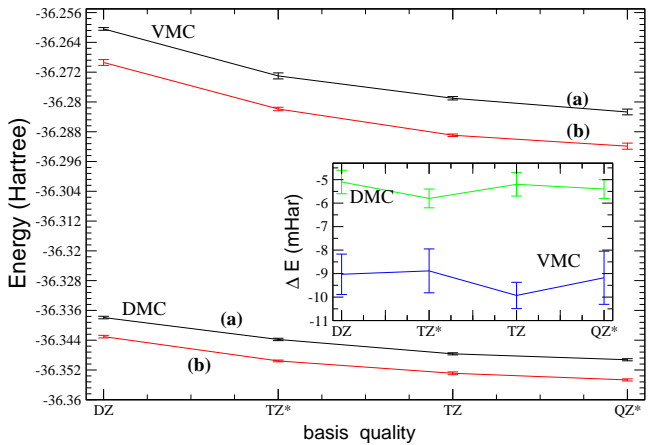}
\caption{VMC and DMC energies for different basis sets obtained with
  a B3LYP trial wave function for the structures \1 and \2. The quality of
  basis sets range from double-zeta (DZ) to quadruple-zeta (QZ) for
  both carbon and hydrogen.  TZ$^*$ is the triple-zeta basis set without
  the carbon $f$-functions, and QZ$^*$ is the quadruple-zeta basis set
  without the carbon $g$-functions. The inset shows $\Delta E = E_b-E_a$
  in mHa.}
\label{fig.basis}
\end{figure}

Figure~\ref{fig.basis} shows the QMC energies obtained, with the
determinant part of the wave function chosen to be a B3LYP single
determinant, for various basis sets.
DZ denotes a double-zeta basis set, TZ$^*$ a
triple-zeta basis set without the $f$-functions of the carbon atoms,
TZ is a triple-zeta basis, and QZ$^*$ denotes a quadruple-zeta basis
without the carbon $g$-functions.
As expected, the VMC and DMC energies improve as the basis is improved.
The DMC energies obtained from  DZ and TZ bases differ by about $8$~milli Hartree (mHa)
and those obtained from TZ and QZ$^*$ bases by less than $2$~mHa.
The $\approx 4$~mHa difference in DMC energy between the TZ and TZ$^*$ basis sets
shows that it helps to include the polarization $f$-functions.

The quantity of interest is the difference in the energies of the
two structures and this converges faster than the individual total energies.
The inset of Fig.~\ref{fig.basis} shows the energy difference $\Delta E = E_b-E_a$
is already converged within statistical errors even with the double-zeta basis.
In the rest of our QMC calculations, we employ only
double- and triple-zeta basis sets.

\subsubsection{CSF convergence}

As seen from Figure~\ref{fig.basis}, when a B3LYP single-determinant
is employed in the wave function, the energy of the \2 structure is lower
than that of the \1 structure.  However, since this is a system where
static correlation is important, it is imperative to employ multi-determinant
wave functions in QMC. In fact inclusion of additional determinants results
in reversing the ordering of the energies.
To this end, we employed truncated complete active space self consistent field (CASSCF)
wave functions to construct the determinant part of our trial wave functions,
which yield lower VMC and DMC energies than the single-determinant wave functions.
The active space was carefully selected such that it contains the six $\pi$
electrons and the two unpaired electrons in structure \1 that form the C1C3
bridging bond in structure \2.

In order to reduce the number of determinants in the QMC calculations,
we include only CSFs which have a coefficient larger in absolute magnitude than a cutoff
parameter, $\eta$, in the MCSCF wave function.  The CSF coefficients
are then reoptimized within VMC in the presence of the Jastrow; 
typically, all the CSF coefficients (except the dominant one) get
smaller in absolute magnitude upon reoptimization
because of the presence of the Jastrow factor.
Table~\ref{tab:csf_convg} shows the QMC
energies for several values of the cutoff $\eta$.
With a cutoff of $\eta = 0.01$, the DMC total energies are converged to better than
$1$~mHa, and this is the value of $\eta$ we used in the rest of our calculations.
Moreover, there is a good degree of cancellation of error between the energies of the
two structures, so that the energy differences are converged, within statistical
errors, even with larger values of $\eta$.

\begin{table*}[tbh]
\caption{
\label{tab:csf_convg}
The convergence of the QMC energies for structures
\1 and \2 with respect to the threshold $\eta$ used to select
the CSFs included in the trial wave function. The determinant part
of the wave function comes from the MCSCF/CAS(8,8) wave function with a double-zeta basis.
Statistical errors on the last digit are shown in parentheses.}
\begin{ruledtabular}
\begin{tabular}{lccccccccc}
& \multicolumn{2}{c}{$E_a$ (Ha)} & &
  \multicolumn{2}{c}{$E_b$ (Ha)} & &
  \multicolumn{2}{c}{$\Delta E=E_b-E_a$ (kcal/mole)}\\
\cline{2-3}  \cline{5-6}  \cline{8-9}
\rm{$\eta$}    & \multicolumn{1}{c}{VMC} & \multicolumn{1}{c}{DMC} & & \multicolumn{1}{c}{VMC} & \multicolumn{1}{c}{DMC} & & \multicolumn{1}{c}{VMC} & \multicolumn{1}{c}{DMC}  \\ \hline
0.05    &   $-$36.2832(5)  & $-$36.3519(3) &  &
$-$36.2757(5)&$-$36.3465(3) & &4.7(4)& 3.4(3)  \\
0.01    &   $-$36.2855(5)  & $-$36.3526(2) &  &  $-$36.2801(5) &
$-$36.3487(2)& &3.4(4)& 2.4(2) \\
0.005   &   $-$36.2869(5)  & $-$36.3531(4) &  &  $-$36.2798(3)
&$-$36.3489(2) & & 4.5(4) &  2.6(3) \\
\end{tabular}
\end{ruledtabular}
\end{table*}

\subsubsection{Time-step convergence}

The imaginary-time evolution operator used in DMC is exact only in the
limit of zero Trotter time step, $\delta\tau$.  Consequently, energies
evaluated at finite $\delta\tau$ have a time-step error which may be
positive or negative.  Table~\ref{tab:trotter_convg} shows the DMC
energies for the two structures and their difference for various
values of $\delta\tau$.  The time-step errors of the energies of the two
structures cancel out within statistical error upon taking the energy
difference.  Consequently, we employed a time step of
$\delta\tau=0.1$~Hartree$^{-1}$ in the rest of our calculations.

\begin{table}[htb]
\caption{
\label{tab:trotter_convg}
DMC energies for different Trotter time steps $\delta\tau$, using a CAS trial wave
function and a double-zeta basis.  Total energies, $E_a, E_b$,
of structures \1 and \2 are in Hartrees, and the energy difference,
$\Delta E = E_b-E_a$, is in kcal/mole.
Statistical errors of the last digit are shown between parenthesis.
$\Delta E$ has a negligible time-step error because the errors in
$E_a$ and $E_b$ nearly cancel.
}
\begin{ruledtabular}
\begin{tabular}{lcccc}
 $\delta\tau$ &  $E_a$ (Ha) & $E_b$ (Ha) & &  $\Delta E$ (kcal/mole)\\
\hline
0.2     & $-$36.3572(3) & $-$36.3533(4) & & 2.4(3) \\
0.1     & $-$36.3526(2) & $-$36.3487(2) & & 2.4(2) \\
0.07    & $-$36.3519(3) & $-$36.3483(3) & & 2.3(3) \\
0.05    & $-$36.3529(3) & $-$36.3487(3) & & 2.6(3)\\
0.01    & $-$36.3563(5) & $-$36.3528(6) & & 2.2(5) \\
\end{tabular}
\end{ruledtabular}
\end{table}

\begin{table}[htb]
\caption{
\label{tab:summary_deltaE}
Energy differences,  $\Delta E = E_b - E_a$ (in kcal/mole)
obtained from BLYP, B3LYP, MCSCF/CAS(8,8) methods, and, from VMC and DMC
using wave functions constructed from the BLYP, B3LYP, MCSCF/CAS(8,8) methods.
For MCSCF both double-zeta (DZ) and triple-zeta (TZ) bases are used.
The geometries are the same as in Table~1.
Statistical errors of the last digit are shown in parentheses.}
\begin{ruledtabular}
\begin{tabular}{lddd}
              & \multicolumn{1}{r}{DFT/MCSCF} &  \multicolumn{1}{r}{VMC} &  \multicolumn{1}{r}{DMC} \\
\hline
B3LYP/TZ  &  -3.75  &   -7.3(4)  &  -3.4(2)  \\
BLYP/TZ   &   1.59  &  -6.2(4)  &  -3.5(2)  \\
MCSCF/CAS(8,8)/DZ &  14.72  &   3.3(4)  &   2.4(2)  \\
MCSCF/CAS(8,8)/TZ &  13.17  &   3.4(5)  &   1.9(2) 
\end{tabular}
\end{ruledtabular}
\end{table}

\section{Results}

Table~\ref{tab:summary_deltaE} shows the energy difference of the two structures
$\Delta E = E_b - E_a$ obtained from the BLYP and B3LYP
density functional methods and from MCSCF/CAS(8,8) theory.
The density functional calculations are converged with respect to
basis size; the difference between the triple- and quadruple-zeta
basis sets being less than $0.1$~kcal/mole. In the CAS(8,8) calculations,
we used double- and triple-zeta basis sets. In agreement with previous
papers, B3LYP favors
structure \2, BLYP favors structure \1, and MCSCF greatly favors
structure \1.

The energy differences $\Delta E$ at the B3LYP and
BLYP levels of theory are in good agreement with the values of $-2.9$ and
$3.0$ kcal/mole obtained using an all electron 6-311++G(3df,3pd) basis set
\cite{Kraka01}. Also, the CAS(8,8) energy difference with
the triple-zeta basis is in very good agreement
with the all-electron value, $12.9$~kcal/mole, obtained using a
6-31+G(2df,2p) basis value at the same level of theory~\cite{Kraka01}. This is a
testament to the good quality of the Burkatski \etal
~\cite{BurkatzkiFilippiDolg07} pseudopotentials and basis sets
employed in our calculations.  The geometries
we used from ~\cite{SmithCrawfordCremer05} are sufficiently close to
those of Ref.~\cite{Kraka01} that they should not appreciably affect
the energy differences.

Table~\ref{tab:summary_deltaE} shows also the QMC energies obtained
using DFT and CAS(8,8) determinants to construct the QMC wave functions.
Note that the QMC energy differences obtained using the two single determinants in the
trial wave functions agree within statistical error, despite the fact
that B3LYP favors structure \2 and BLYP favors structure \1.
It appears that it is necessary for this system to employ multi-determinant wave functions
to obtain the correct energy ordering in QMC.

The best DMC result is obtained with the CAS(8,8) wave function;
this wave function captures the static correlations that are important for this system and
has a lower energy and lower root-mean-square fluctuations of the 
energy than the single determinant wave functions.
We show in Table~\ref{tab:summary_deltaE} the values obtained with the
double- and triple-zeta basis sets; there is no dependence on the
basis size within statistical error.  Our best estimate is that
structure \1 has a lower energy than structure \2 by 1.9(2) kcal/mole.
This result is in excellent agreement with the best existing quantum
chemistry calculations, namely $1.0$~kcal/mole from CCSD(T) using a
6-311++G(3df,3pd) basis~\cite{Kraka01} and $\approx 1.6$~kcal/mole
from CCSD(T) using a cc-pVTZ basis~\cite{SmithCrawfordCremer05}.  The
result is also in reasonable agreement with the $4.7$~kcal/mole energy
difference obtained using a CASSCF(8,8)/6-31+G(2df,2p) wave function
followed by second order perturbation theory (CASPT2)~\cite{Kraka01}.

In summary, our DMC results support the CCSD(T) and CCSDT conclusions
\cite{Kraka01,SmithCrawfordCremer05} that the cyclic structure \1 is
the ground state of the system. We find that accurate fixed-node DMC energy differences
for the structures of m-benzyne require multi-determinant wave functions;
the fixed-node error in the energy difference with a single determinant from
restricted B3LYP or BLYP calculations is $\approx 5$~kcal/mole.
Our best DMC estimate, obtained with a reoptimized MCSCF/CAS(8,8) trial wave function,
shows that the proposed monocyclic structure of m-benzyne is lower in energy than the bicyclic
structure by $1.9(2)$~kcal/mole, in very good
agreement with previous coupled cluster estimates.

\section{Acknowledgments}
We thank Julien Toulouse for valuable discussions, and
Claudia Filippi for providing pseudopotentials prior to publication.
This work was supported in part by the National Science Foundation (EAR-0530301) and the DOE (DE-FG02-07ER46365).
The calculations were performed on the Intel cluster at the Cornell Nanoscale Facility
(a member of the National Nanotechnology Infrastructure Network supported by the NSF)
and at the Cornell Theory Center.

%\bibliographystyle{apsrev}

%\bibliography{qmc}

\end{document}